\documentclass[showpacs]{revtex4}
\usepackage{amsmath}
\usepackage{amssymb}
\usepackage{graphicx}
\usepackage{color}
\usepackage{dcolumn}

\begin{document}

\title{
Hiding a neutron star inside a wormhole
}
\author{
Vladimir Dzhunushaliev,$^{1,2,3,4}$
\footnote{v.dzhunushaliev@gmail.com}
Vladimir Folomeev,$^{2,3}$
\footnote{vfolomeev@mail.ru}
Burkhard Kleihaus,$^{4}$
\footnote{b.kleihaus@uni-oldenburg.de }
Jutta Kunz$^4$
\footnote{jutta.kunz@uni-oldenburg.de}
}
\affiliation{$^1$
Department of Theoretical and Nuclear Physics,  Al-Farabi Kazakh National University, Almaty 050040, Kazakhstan \\
$^2$ Institute of Experimental and Theoretical Physics,  Al-Farabi Kazakh National University, Almaty 050040, Kazakhstan \\
$^3$Institute of Physicotechnical Problems and Material Science of the NAS
of the Kyrgyz Republic, 265 a, Chui Street, Bishkek 720071,  Kyrgyz Republic \\
$^4$Institut f\"ur Physik, Universit\"at Oldenburg, Postfach 2503
D-26111 Oldenburg, Germany
}

\begin{abstract}
We consider neutron-star-plus-wormhole configurations supported by a massless ghost scalar field.
The neutron fluid is modeled by an anisotropic equation of state.
When the central energy density of the fluid is of comparable magnitude
to the one of the scalar field, configurations
with an equator at the center and a double throat arise.
These double-throat wormholes can be either partially or completely filled
by the neutron fluid.
In the latter case, the passage of light -- radiated by the neutron matter --
through these wormholes is studied.
A stability analysis indicates that all considered configurations are unstable
with respect to linear perturbations, independent of whether the fluid is isotropic or anisotropic.
\end{abstract}

\maketitle

\section{Introduction}

At the present time it is widely believed that the main contributions
to the total energy density of the Universe come from
dark energy (DE), dark matter (DM), and ordinary matter.
Distributed homogeneously over all of space,
DE contributes about  70\% of the total energy density.
Since DE possesses a large negative pressure, it
drives the present accelerated expansion of the Universe.
In turn, DM contributes about 25\% of the total energy density.
DM is gravitationally clustered in galaxies and galaxy clusters
and responsible for  the formation of the large-scale structure of the Universe.
Finally, ordinary (baryonic)  matter forms the visible matter
and represents about 5\% of the total energy density of the Universe.

The main feature of DE and DM is their extremely weak ability
to participate in the electromagnetic interaction,
which hampers considerably  their direct observation.
However, like ordinary matter, these two substances take part
in the gravitational interaction, possibly leading to the creation of
localized compact objects composed of these substances.
Indeed various models of compact objects
consisting of dark energy \cite{de_stars}, of dark matter
\cite{dm_stars}, or of
interacting DE and DM \cite{de_dm_conf} are considered in the literature.

The unusual physical properties of DE,
which enable us to model the accelerated expansion of the present Universe,
may lead to other interesting consequences as well.
In particular, when considering compact objects composed of DE
it is possible to obtain two essentially
different types of systems -- exhibiting either a trivial or a nontrivial topology of spacetime.
The  so-called dark energy stars \cite{de_stars}, which are modeled by some
form of matter possessing the properties of DE,
belong to the first type.

For these configurations the strong energy condition is violated
when the effective pressure $p$ of such matter
satisfies the inequality $p< -\varepsilon/3$,
where $\varepsilon$ is the effective energy density.
When $p< -\varepsilon$, even the null energy condition is violated.
The matter is then called exotic.
The presence of exotic matter allows for compact configurations with a nontrivial
wormholelike topology.
In the simplest case such configurations can be supported
by the so-called ghost (or phantom) scalar fields,
which may be massless~\cite{wh_massless} or possess a potential energy~\cite{wh_quart}.
(For further discussion of different aspects of phantom field wormholes see, e.g.,
Refs.~\cite{Thorne:1988,kuhfittig,lxli,ArmendarizPicon:2002km,Sushkov:2002ef,Lemos:2003jb,Lobo:2005us,Sushkov:2005kj},
and a  general overview on the subject of Lorentzian wormholes can be found in the book~\cite{Visser}.)
Ghost (or phantom) scalar fields also allow for ``black universes,'' i.e., black holes with
an expanding universe inside the horizon
\cite{Bronnikov:2005gm,Bronnikov:2006fu,Bolokhov:2012kn,Bronnikov:2012ch}.

Another possibility is to consider mixed configurations with nontrivial topology
which consist both of ordinary and exotic matter.
In Refs.~\cite{Dzhunushaliev:2011xx,Dzhunushaliev:2012ke,Dzhunushaliev:2013lna}
we have studied such mixed compact configurations, where a wormhole
(supported by a ghost scalar field) is filled by neutron matter.
The resulting neutron-star-plus-wormhole configurations
then possess properties of wormholes and of ordinary stars.
In particular, as discussed in Ref.~\cite{Dzhunushaliev:2013lna},
such systems have an important dimensionless parameter called $B$,
which corresponds to the ratio of the neutron matter energy density to that
of the scalar field at the center.
In Ref.~\cite{Dzhunushaliev:2013lna} we have considered configurations with
$B\ll 1$ for the case of a quartic potential of the scalar field.

Here we study the case with  $B\sim 1$ and show that this can lead to interesting consequences.
In particular, we show that the neutron-star-plus-wormhole configuration
need no longer possess a single throat located at the center.
Instead, an equator may arise at the center with two throats formed
away from the center, each associated with an asymptotically flat universe.
The neutron matter may then fill the wormhole beyond both throats,
or it may be completely confined within the space between the throats.

Let us here briefly comment on the terminology used in the following.
Ordinary stars with trivial topology
possess a true center, located at the point
with radial coordinate $r=0$. In contrast,
for the configurations with nontrivial topology considered here,
the value $r=0$ does not describe a center in the usual sense.
Indeed, at $r=0$ the radius of a two-sphere does not vanish,
but assumes  a (local) nonzero extremum:
a minimum in the case of a throat or a maximum in the case of an equator.
Previously, in Refs.~\cite{Dzhunushaliev:2012ke,Dzhunushaliev:2013lna}
we employed the term ``core'' for the region around the throat.
But since for the configurations considered here
besides a throat also an equator may reside at $r=0$,
we will use the term ``center''  when referring to $r=0$.
The term center thus refers to the extremal surface,
located symmetrically between the two asymptotically flat regions.

To construct the wormhole, we employ a massless ghost scalar field,
and for the ordinary matter we take a neutron fluid with an anisotropic equation of state (EOS).
Such an EOS assumes that
massive stellar objects may have a radial pressure
that is not equal to the tangential pressure
at high densities of the neutron matter.
There are several physical reasons for the appearance of an anisotropy
(see Ref.~\cite{anis_reas} for some of them).
Thus one should take into account the effects of an anisotropy, in particular,
in modeling the solid cores of neutron stars \cite{Glendenning:1997}.
In turn, the presence of an anisotropy of the fluid results in
considerable changes of the characteristics of relativistic stars,
as considered, for example, in Refs.~\cite{Bowers:1974,Heintzmann:1975,Hillebrandt:1976,Bayin:1982,Herrera:1997,Mak:2001eb}.
Our goal here is to clarify the question of how the presence of an anisotropy of the fluid
influences the properties of the mixed configurations under consideration,
e.g., their masses and sizes,
and their stability with respect to linear perturbations.

The paper is organized as follows.
In Sec.~\ref{statem_prob} the statement of the problem is presented.
Here we describe the matter components
appearing in the system and write down the corresponding field equations.
In Sec.~\ref{num_calc} we construct explicit examples of regular static solutions
describing neutron-star-plus-wormhole configurations.
We evaluate their masses and sizes, consider the question of light
passing through the wormhole,
and estimate the tidal effects in the system.
In Sec.~\ref{linear_stab} a linear stability analysis is performed for these solutions.
Finally, in Sec.~\ref{conclusion} we summarize the results obtained.

\section{Statement of the problem}
\label{statem_prob}

Here we consider a gravitating system consisting of a wormhole  supported by a ghost scalar field $\varphi$ and filled by
ordinary matter in the form of a neutron fluid. For simplicity we consider a massless scalar field.
We know that in the presence of a scalar field the effective pressure becomes anisotropic
(see, e.g., Ref.~\cite{Gleiser:1988rq} and references therein).
Apart from this anisotropy associated with the scalar field,
we here assume that the neutron matter also possesses an anisotropic pressure,
where the radial and tangential components of the pressure are not equal to each other.

The Lagrangian of this system can be chosen as follows:
\begin{equation}
\label{lagran_wh_star_poten}
L=-\frac{c^4}{16\pi G}R+L_{\text{sf}}+L_{\text{fl}},
\end{equation}
with the curvature scalar $R$ and
Newton's constant $G$,
the Lagrangian of the ghost scalar field $L_{\text{sf}}$,
\begin{equation}
\label{lagran_sf}
L_{\text{sf}}=-\frac{1}{2}\partial_{i}\varphi\partial^{i}\varphi
 , \end{equation}
and the Lagrangian of the fluid $L_{\text{fl}}$.
Both matter sources contribute
to the right-hand side of the Einstein equations described below.

\subsection{Anisotropic fluid}
\label{anis_fluid}

Since at the moment  there exists no reliable theory modeling the anisotropy of
a neutron fluid at high densities,
here we employ one of the approaches of Refs.~\cite{Heintzmann:1975,Hillebrandt:1976}.
In this case the
energy-momentum tensor of the neutron matter is chosen in the form
\begin{equation}
\label{fluid_emt}
T_{i (\text{fl})}^{k}=(\varepsilon^*+p)u_i u^k-\delta_i^k p+\Theta_i^k.
\end{equation}
Here $\varepsilon^*$, $p$, and $u^i$ are the energy density, the pressure,
and the four-velocity of the fluid, respectively,
and $\Theta_i^k$ is a trace-free ``shear-stress'' tensor,
that assumes $\Theta_\mu^\mu=0$, and $p=1/3\, p_\mu^\mu$ (here the Latin indices
run over $i,k,...=0,1,2,3$ and the Greek indices $\mu, \nu...= 1,2,3$). Following Refs.~\cite{Heintzmann:1975,Hillebrandt:1976},
we assume $\Theta_i^k$ to be diagonal,
and $\Theta_0^0+\varepsilon^*\equiv \varepsilon$ to be the total energy density of the fluid.
Also, we assume that $\Theta^\mu_\nu$ has only one independent component $\Theta_1^1=\alpha p$, while the other two
components can be expressed in terms of $\Theta_1^1$,
$\Theta_2^2=\Theta_3^3=-(\alpha/2) p$ \cite{Heintzmann:1975,Hillebrandt:1976}.
In general,
$\alpha$ can be a function of the matter variables and of the coordinates, $\alpha=\alpha(\varepsilon, x^i)$.

With this  ansatz, we have the following components of the fluid energy-momentum tensor:
\begin{equation}
\label{fluid_emt_comp}
T_{0 (\text{fl})}^0=\varepsilon, \quad T_{1 (\text{fl})}^1=-p_r, \quad T_{2 (\text{fl})}^2=T_{3 (\text{fl})}^3=-p_t,
\end{equation}
where the radial and the tangential components of the pressure are
\begin{equation}
\label{fluid_pressures}
p_r=(1-\alpha)p,  \quad p_t=(1+\alpha/2)p.
\end{equation}
(These expressions show that one must choose $-2<\alpha<1$
in order to have a positive fluid pressure.)
With these expressions one can eliminate $p$,
which yields the following relation between the pressure components:
\begin{equation}
\label{parameter_beta}
p_t=(1+\beta)p_r, \quad
\beta=\frac{3}{2}\frac{\alpha}{1-\alpha}.
\end{equation}
Thus $\beta$ (or equivalently $\alpha$) is the parameter determining the anisotropy
of the fluid (i.e., the anisotropy parameter).

The energy density $\varepsilon$ and the pressure $p$ appearing in the expressions \eqref{fluid_emt_comp} and \eqref{fluid_pressures}
are related by an equation of state that is determined by
the physical properties of the specific matter considered
and by the physical conditions under which it is employed.
Since we here consider essentially relativistic objects,
it is natural to assume that the matter filling
the wormholes should also be relativistic.
Thus we choose relativistic neutron matter for this kind of matter.
In much of the literature such matter is described
by more or less conventional equations of state,
reflecting its general properties at high densities and pressures.
Various forms of such equations of state can be found, for instance,
in Refs.~\cite{Oppen1939,Cameron1959,DAl1985,Haensel:2004nu}.

For our purpose,  we restrict ourselves to a simplified variant of the EOS,
where a more or less realistic neutron matter
EOS is approximated in the form of a polytropic EOS.
This EOS can be taken in the following form:
\begin{equation}
\label{eqs_NS_WH}
p=K \rho_{b}^{1+1/n}, \quad \varepsilon = \rho_b c^2 +n p,
\end{equation}
with the constant $K=k c^2 (n_{b}^{(ch)} m_b)^{1-\gamma}$,
the polytropic index $n=1/(\gamma-1)$,
and $\rho_b=n_{b} m_b$ denotes the rest-mass density
of the neutron fluid. Here $n_{b}$ is the baryon number density,
$n_{b}^{(ch)}$ is a characteristic value of $n_{b}$,
$m_b$ is the baryon mass,
and $k$ and $\gamma$ are parameters
whose values depend on the properties of the neutron matter.

As in our previous works concerning mixed star-plus-wormhole systems
\cite{Dzhunushaliev:2012ke,Dzhunushaliev:2013lna},
we here, for simplicity, take only one set of parameters for the neutron fluid.
Namely, we choose
$m_b=1.66 \times 10^{-24}\, \text{g}$,
$n_{b}^{(ch)} = 0.1\, \text{fm}^{-3}$,
$k=0.1$, and $\gamma=2$ \cite{Salg1994}.
These parameters correspond to a gas of baryons interacting
via a vector-meson field, as described by Zel'dovich \cite{Zeld1961,Zeld}.
We employ these values for the parameters in the numerical calculations of Sec.~\ref{num_calc}.

\subsection{Field equations}

For spherically symmetric systems, the metric  can be taken in the general form \cite{Land1}
 \begin{equation}
\label{metric_gen}
ds^2=e^{\nu}(dx^0)^2-e^{\lambda} dr^2-e^{\mu} d\Omega^2,
\end{equation}
where $\nu$, $\lambda$, and $\mu$ are functions of the radial coordinate $r$
and the time coordinate  $x^0=c\, t$,
and $d\Omega^2$ is the metric on the unit two-sphere. We will use this metric below when considering the question of
linear stability in Sec.~\ref{linear_stab}.

For the construction of equilibrium neutron-star-plus-wormhole configurations
it is convenient  to use polar Gaussian coordinates. The metric then reads
 \begin{equation}
\label{metric_wh_poten}
ds^2=e^{\nu}(dx^0)^2-dr^2-e^{\mu} d\Omega^2,
\end{equation}
where now $\nu$ and $\mu$ are functions of $r$ only.
Introducing the new function $R$ defined by $e^{\mu}=R^2$
and taking into account the components of the energy-momentum tensor of the fluid \eqref{fluid_emt_comp} and
\eqref{fluid_pressures},
the $(_0^0)$, $(_1^1)$, and $(_2^2)$ components of the Einstein equations with the metric
 \eqref{metric_wh_poten}  take the form
\begin{eqnarray}
\label{Einstein-00_poten}
&&-\left[2\frac{R^{\prime\prime}}{R}+\left(\frac{R^\prime}{R}\right)^2\right]+\frac{1}{R^2}
=\frac{8\pi G}{c^4} T_0^0=
\frac{8\pi G}{c^4}\left[ \varepsilon-\frac{1}{2}\varphi^{\prime 2}\right],
 \\
\label{Einstein-11_poten}
&&-\frac{R^\prime}{R}\left(\frac{R^\prime}{R}+\nu^\prime\right)+\frac{1}{R^2}
=\frac{8\pi G}{c^4} T_1^1=
\frac{8\pi G}{c^4}\left[- (1-\alpha)p+\frac{1}{2}\varphi^{\prime 2}\right],
\\
\label{Einstein-22_poten}
&&\frac{R^{\prime\prime}}{R}+\frac{1}{2}\frac{R^\prime}{R}\nu^\prime+
\frac{1}{2}\nu^{\prime\prime}+\frac{1}{4}\nu^{\prime 2}
=-\frac{8\pi G}{c^4} T_2^2=
\frac{8\pi G}{c^4}\left[ \left(1+\frac{\alpha}{2}\right)p+\frac{1}{2}\varphi^{\prime 2}\right],
\end{eqnarray}
where the prime denotes  differentiation with respect to $r$.
Here the corresponding components of the scalar field energy-momentum tensor
have been obtained by varying the Lagrangian \eqref{lagran_sf} with respect to the metric.

The field equation for the scalar field
is obtained by varying the Lagrangian
\eqref{lagran_sf} with respect to $\varphi$:
\begin{equation}
\label{sf_eq_gen}
\frac{1}{\sqrt{-g}}\frac{\partial}{\partial x^\mu}
\left[\sqrt{-g}g^{\mu\nu}\frac{\partial \varphi}{\partial x^\nu}\right]= 0.
\end{equation}
Using the metric
 \eqref{metric_wh_poten}, this equation is integrated to give
\begin{equation}
\label{sf_poten}
\varphi^{\prime 2}=\frac{D^2}{R^4}e^{-\nu},
\end{equation}
where $D$ is an integration constant.

Because of the conservation of energy and momentum,
$T^k_{i; k}=0$,
not all of the Einstein field equations are independent.
Taking the $i=1$ component of the conservation equations gives
\begin{equation}
\label{conserv_1}
\frac{d T^1_1}{d r}+
\frac{1}{2}\left(T_1^1-T_0^0\right)\nu^\prime+2\frac{R^\prime}{R}\left[T_1^1-\frac{1}{2}\left(T^2_2+T^3_3\right)\right]=0.
\end{equation}
Taking into account the expressions for the components
$T_0^0$, $T_1^1$, and $T_2^2=T_3^3$ [see the right-hand sides of Eqs.~\eqref{Einstein-00_poten}-\eqref{Einstein-22_poten}],
and also Eq.~\eqref{sf_poten}, we obtain from Eq.~\eqref{conserv_1}
\begin{equation}
\label{conserv_2}
(1-\alpha)\frac{d p}{d r}+\frac{1}{2}\left[\varepsilon+(1-\alpha)p\,\right]\frac{d\nu}{d r}-3\alpha \frac{R^\prime}{R}p=0.
\end{equation}
Thus we have four unknown functions~-- $R$, $\nu$, $p$, and $\varphi$~-- for which there are five equations,
\eqref{Einstein-00_poten}-\eqref{Einstein-22_poten},
\eqref{sf_poten}, and \eqref{conserv_2}, only four of which are independent.

For the numerical calculations it is convenient to rewrite
these equations in terms of dimensionless variables.
For the case considered here the massless scalar field $\varphi$
can be taken,  without loss of generality, equal to zero
at the center of the configuration, i.e., at $r=0$,
while its derivative at $r=0$ is nonzero.
The potential of the scalar field can be expanded  in the neighborhood of the center as
$$
\varphi \approx \varphi_1 r +\frac{1}{6}\varphi_3 r^3,
$$
where $\varphi_1$ is the derivative at the center,
the square of which corresponds to the ``kinetic'' energy of the scalar field.
Then it is convenient to use new dimensionless variables
expressed in units of $\varphi_1$.
Namely, introducing
\begin{equation}
\label{dimless_xi_v}
\xi=\frac{r}{L}, \quad \Sigma=\frac{R}{L},
\quad \phi(\xi)=\frac{\sqrt{8\pi G}}{c^2}\,\varphi(r),
\quad \text{where} \quad L=\frac{c^2}{\sqrt{8\pi G}\varphi_1},
\end{equation}
and using the new reparametrization of the fluid density~\cite{Zeld},
\begin{equation}
\label{theta_def}
\rho_b=\rho_{b c} \theta^n~,
\end{equation}
where $\rho_{b c}$ is the density of the neutron fluid at the
center of the configuration, we rewrite Eqs.~\eqref{Einstein-00_poten}-\eqref{Einstein-22_poten},
\eqref{sf_poten}, and \eqref{conserv_2} in the dimensionless form
\begin{eqnarray}
\label{Einstein-00_dmls}
&&-\left[2\frac{\Sigma^{\prime\prime}}{\Sigma}+\left(\frac{\Sigma^\prime}{\Sigma}\right)^2\right]+\frac{1}{\Sigma^2}
=B  (1+\sigma n \theta) \theta^n
-\frac{1}{2}\phi^{\prime 2},
 \\
\label{Einstein-11_dmls}
&&-\frac{\Sigma^\prime}{\Sigma}\left(\frac{\Sigma^\prime}{\Sigma}+\nu^\prime\right)+\frac{1}{\Sigma^2}
=-B  \sigma (1-\alpha) \theta^{n+1}
+\frac{1}{2}\phi^{\prime 2},
\\
\label{Einstein-22_dmls}
&&\frac{\Sigma^{\prime\prime}}{\Sigma}+\frac{1}{2}\frac{\Sigma^\prime}{\Sigma}\nu^\prime+
\frac{1}{2}\nu^{\prime\prime}+\frac{1}{4}\nu^{\prime 2}
=
B \sigma \left(1+\frac{\alpha}{2}\right) \theta^{n+1}
+\frac{1}{2}\phi^{\prime 2},
\\
\label{sf_dmls}
&&\phi^{\prime 2}=\frac{e^{\nu_c-\nu}}{(\Sigma/\Sigma_c)^4},
\\
\label{fluid_dmls}
&&\sigma(n+1)(1-\alpha)\theta^\prime+\frac{1}{2}\left[1+\sigma(n+1-\alpha)\theta\right]\nu^\prime
-3\alpha \sigma\theta\frac{\Sigma^\prime}{\Sigma}=0.
\end{eqnarray}
Here
$B=(\rho_{b c} c^2)/\varphi_1^2$ is the dimensionless ratio of the fluid energy density to that
of the scalar field at the center; $\Sigma_c$ and $\nu_c$ are the central values of the corresponding functions
[see Eq.~\eqref{bound_stat}];
the integration constant from \eqref{sf_poten} is chosen as $D^2=\left(c^4/8\pi G \varphi_1\right)^2 \Sigma_c^4 e^{\nu_c}$
to provide $\phi^\prime=1$ at the center; $\sigma=K \rho_{b c}^{1/n}/c^2=p_c/(\rho_{b c} c^2)$ is a constant,
related to the pressure $p_c$ of the fluid at the center. The values of the fluid parameters appearing here
 are taken from the end of Sec.~\ref{anis_fluid}.

\subsection{Boundary conditions}

We here consider neutron-star-plus-wormhole configurations
that are asymptotically flat and
symmetric under $\xi \to -\xi$.
The metric function $\Sigma(\xi)$ may be considered
as a dimensionless circumferential radial coordinate.
Asymptotic flatness requires that $\Sigma(\xi) \to |\xi|$ for large $|\xi|$.
Because of the assumed symmetry of the configurations,
the center of the configurations at
$\xi=0$ should correspond to an extremum
of $\Sigma(\xi)$, i.e.,~$\Sigma'(0)=0$.
If $\Sigma(\xi)$ has a minimum at $\xi=0$, then $\xi=0$ corresponds to the throat
of the wormhole.
If, on the other hand, $\Sigma(\xi)$ has a local maximum at $\xi=0$,
then $\xi=0$ corresponds to an equator.
In that case, the wormhole will have a double throat
surrounding a belly (see, e.g., Refs.~\cite{Charalampidis:2013ixa,Hauser:2013jea}).

Expanding the metric function
$\Sigma$ in the neighborhood of the center
$$\Sigma\approx \Sigma_c+1/2\, \Sigma_2 \xi^2$$
and using Eqs.~\eqref{Einstein-00_dmls} and \eqref{Einstein-11_dmls}, we find
the relations
\begin{equation}
\label{bound_coef}
\Sigma_c=\frac{1}{\sqrt{1/2-B \sigma(1-\alpha)}}, \quad \Sigma_2=\frac{\Sigma_c}{2}\Big\{1-B\left[1+\sigma(n+1-\alpha)\right]\Big\}.
\end{equation}
Thus the sign of the expansion coefficient $\Sigma_2$ determines whether
the configurations possess a single throat at the center or an equator
surrounded by a double throat.

Equations \eqref{Einstein-00_dmls}-\eqref{fluid_dmls} are solved
for given parameters of the fluid $\sigma$, $n$, and $B$,
subject to the boundary conditions at the center of the configuration $\xi=0$,
\begin{equation}
\label{bound_stat}
\theta(0) = 1, \quad \Sigma(0)=\Sigma_c, \quad \Sigma^\prime (0)=0,
\quad \nu(0)=\nu_c, \quad
\phi(0) =0, \quad \phi^\prime (0)=1.
\end{equation}
Note here that, using \eqref{dimless_xi_v}, we can express the dimensional value of the derivative $\varphi_1$  as follows:
$$
\varphi_1^2=\frac{c^4}{8\pi G}\frac{1}{L^2}.
$$
Thus the dimensional ``kinetic'' energy of the scalar field depends only on the value of the characteristic length $L$,
which can be chosen arbitrarily subject to some physically reasonable assumptions.
Substituting this $\varphi_1^2$ into
the expression for $B$ [see  Eq.~\eqref{fluid_dmls} below], we find
$$
B=8\pi G \rho_{b c} (L/c)^2.
$$
It is seen
from the above expressions for $\varphi_1^2$ and $B$ that by fixing $L$,
one automatically determines the value of $\varphi_1^2$.
But the value of $B$ can still change depending on
the central value of the fluid density $\rho_{b c}$.
Therefore one can consider $B$ as a parameter
describing the ratio of the fluid energy density at the center
to the energy density of the scalar field at the center.

\section{Numerical results}
\label{num_calc}

In Ref.~\cite{Dzhunushaliev:2013lna} we studied mixed neutron-star-plus-wormhole systems
supported by a ghost scalar field with a quartic potential,
restricting our investigations to small values of $B$, $B\lesssim 0.1$.
Here our aim is to study such mixed configurations, in particular, also for large values of $B$.
Moreover, we here study the effect of anisotropy.

We solve the system of equations \eqref{Einstein-00_dmls}-\eqref{fluid_dmls}
numerically using the boundary conditions \eqref{bound_coef} and \eqref{bound_stat}.
In doing so, the configurations under consideration can be subdivided into two regions:
 (i) the internal one, where both the scalar field and the fluid are present;
 (ii) the external one, where only the scalar field is present.
Correspondingly, the solutions in the external region are obtained by using Eqs.~\eqref{Einstein-00_dmls}-\eqref{sf_dmls},
in which $\theta$ is set to zero.

The internal solutions must be matched with the external ones at the boundary of the fluid,
$\xi=\xi_b$, by equating the corresponding values of the functions $\phi$, $\Sigma$, $\nu$
and their derivatives.
The boundary of the fluid $\xi_b$ is defined by $p(\xi_b)=0$.
Knowledge of the asymptotic solutions in turn allows one
to determine the value of the integration constant $\nu_c$ at the center,
proceeding from the requirement
of asymptotic flatness of the external solutions.

\begin{figure}[t]
\begin{minipage}[t]{1\linewidth}
  \begin{center}
  \includegraphics[height=7cm]{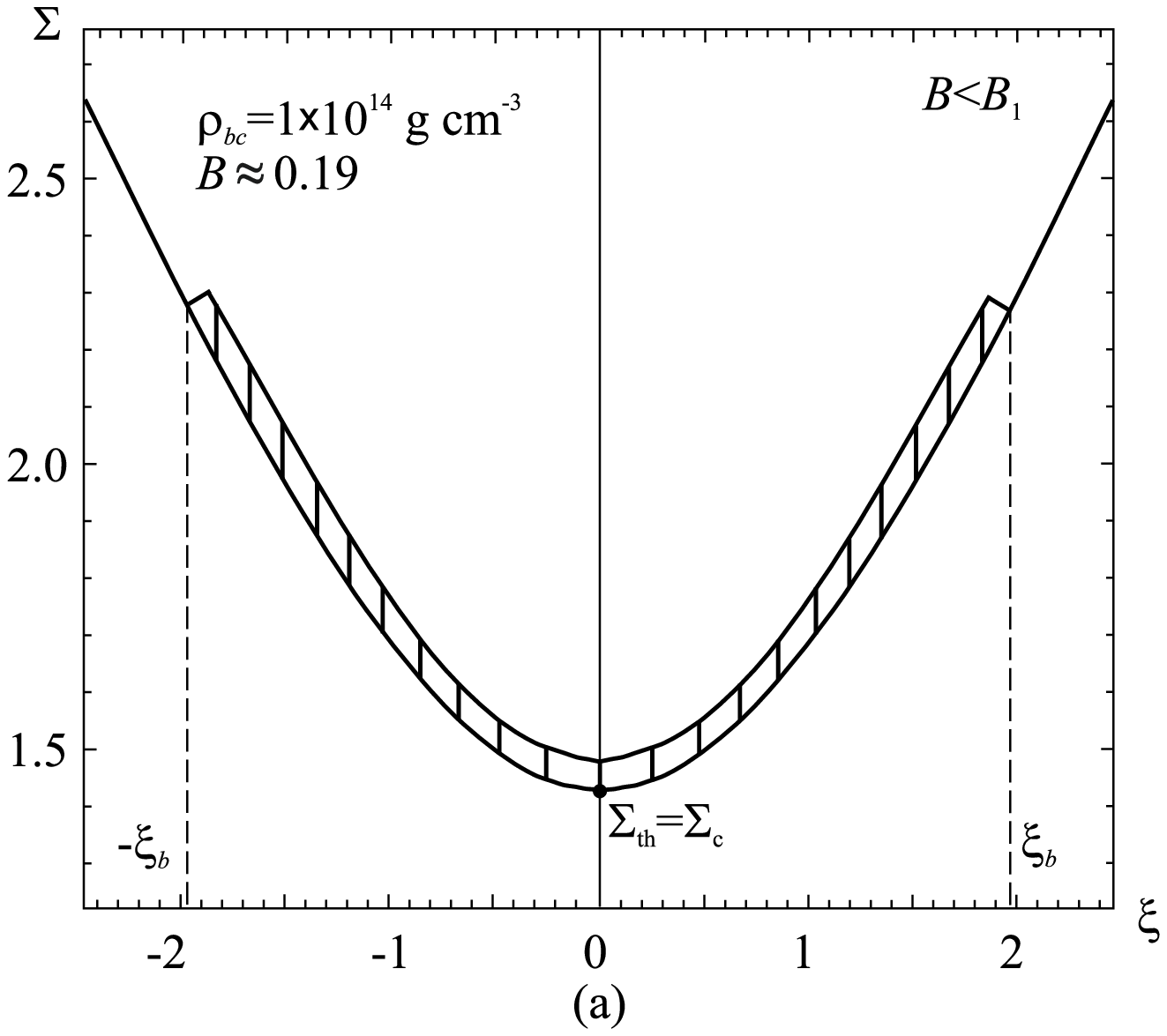}
  \end{center}
\end{minipage}\hfill
\begin{minipage}[t]{.49\linewidth}
  \begin{center}
  \includegraphics[width=7cm]{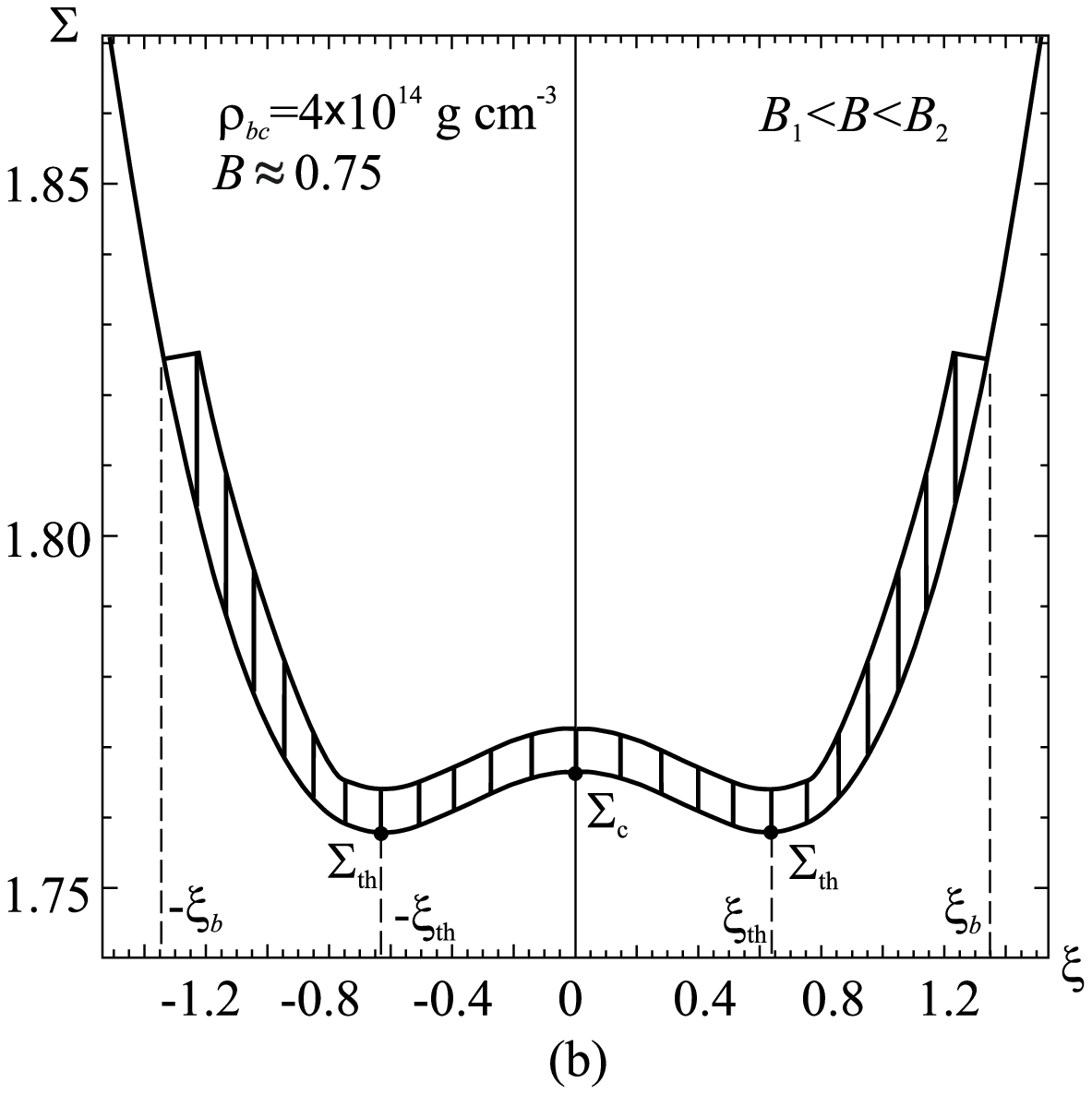}
  \end{center}
\end{minipage}\hfill
\begin{minipage}[t]{.49\linewidth}
  \begin{center}
  \includegraphics[width=7cm]{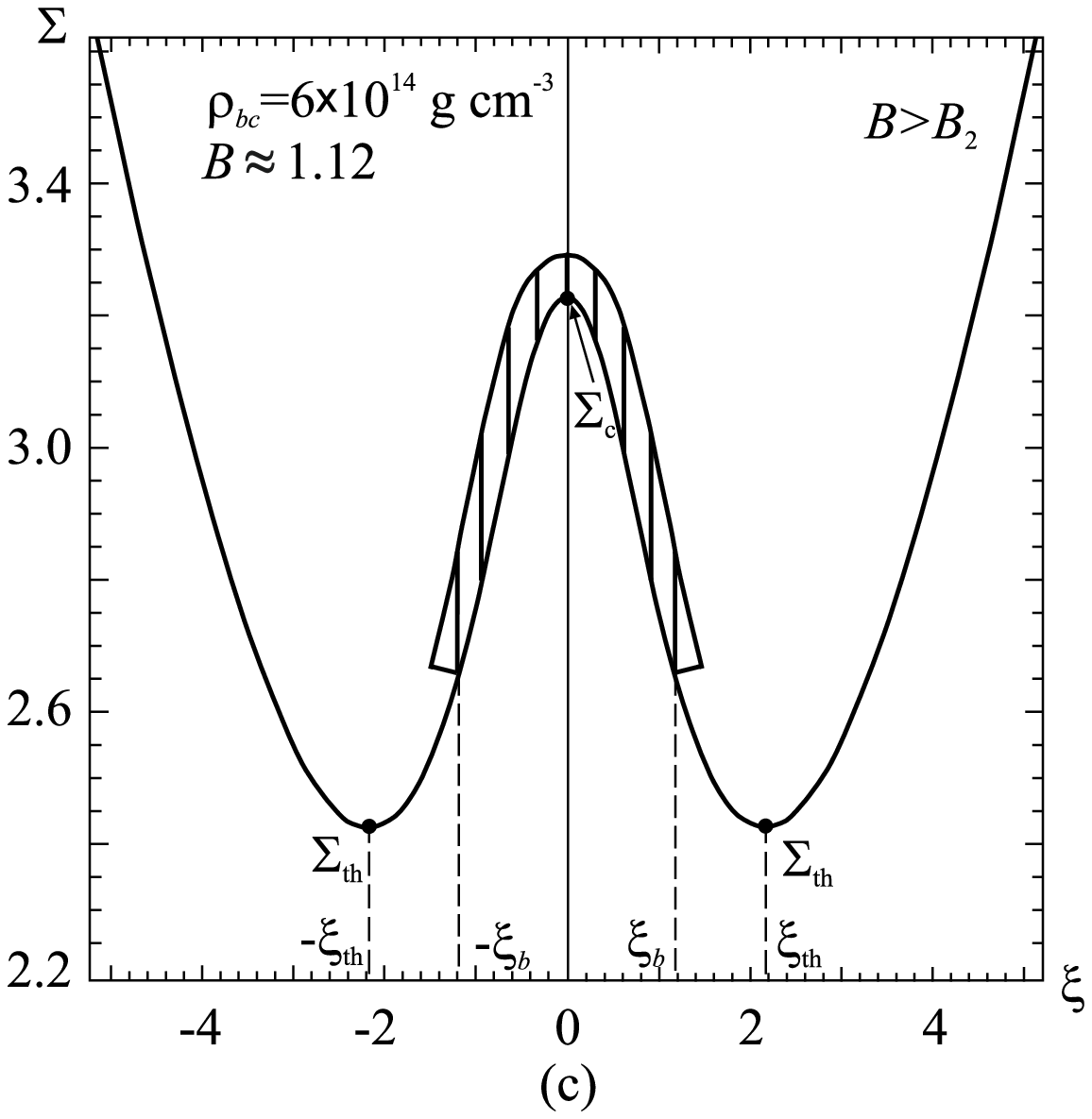}
  \end{center}
\end{minipage}\hfill
\caption{Examples of positions of the throat(s)
$\Sigma_{\text{th}}$ depending on the value of the parameter $B$
(or equivalently $\rho_{bc}$). The shaded areas represent the regions where the fluid is present.
Here $\xi_{\text{th}}$ and $\xi_b$ correspond to the positions of the throat(s)
and the boundary of the fluid, respectively.
For all plots, the characteristic size $L$ is taken as 10 km, and
the value of the anisotropy parameter $\alpha$ is taken to be zero.
}
\label{fig_sigma}
\end{figure}

Even without solving the equations, we see already from \eqref{bound_coef} that
there exists a critical value of $B$, $B_{\text{crit}}$, at which $\Sigma_c\to \infty$.
Physical solutions exist only for $B<B_{\text{crit}}$.
In fact, we find three possible types of solutions:
\begin{enumerate}
\item[(1)]  For small values of $B$, where $\Sigma_2>0$, there is a single throat,
located at the center of the configuration. Its size is
$\Sigma_{\text{th}}=\Sigma_c$ [see Fig.~\ref{fig_sigma}(a)].
\item[(2)] For increasing values of $B$ a particular value $B_1<B_{\text{crit}}$ is encountered,
where $\Sigma_2$ changes sign. For $B>B_1$ the center of the configuration
no longer represents a throat but instead corresponds to an equator.
On each side of the equator a minimal area surface and thus a throat is located,
i.e., $\Sigma_{\text{th}} < \Sigma_c$.
The resulting configurations represent double-throat systems,
where the throats are still filled by the fluid; see Fig.~\ref{fig_sigma}(b).
\item[(3)] Finally, for still larger values of $B$ there exists another special value $B_2$,
that lies in the range $B_1<B_2<B_{\text{crit}}$,
where the fluid just reaches up to the throats.
For $B>B_2$ the throats are then located beyond the fluid,
i.e., the fluid is completely hidden in the belly region between the throats
[see Fig.~\ref{fig_sigma}(c)].
\end{enumerate}

\begin{figure}[t]
\begin{minipage}[t]{.49\linewidth}
  \begin{center}
  \includegraphics[width=7cm]{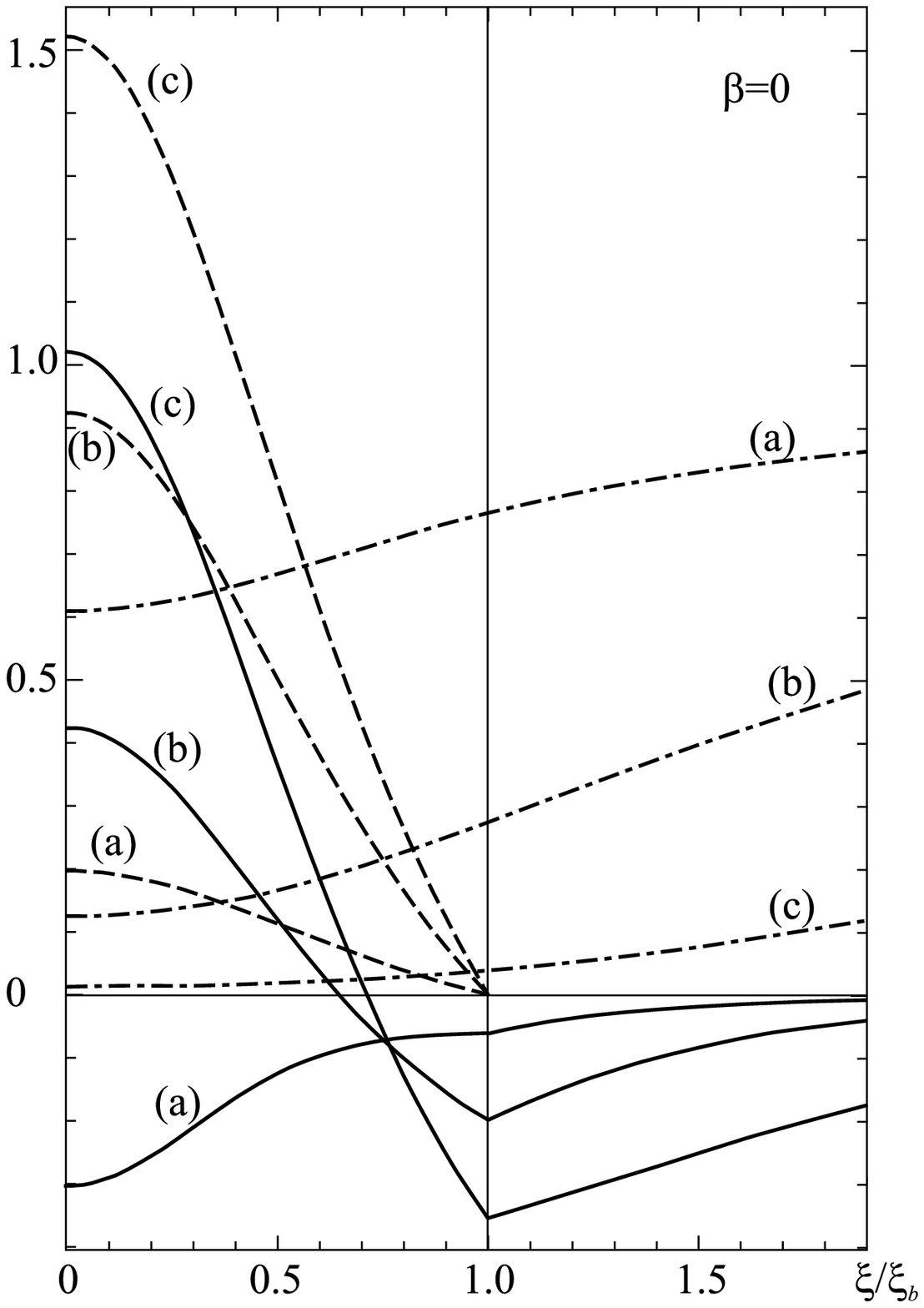}
  \end{center}
\end{minipage}\hfill
\begin{minipage}[t]{.49\linewidth}
  \begin{center}
  \includegraphics[width=7cm]{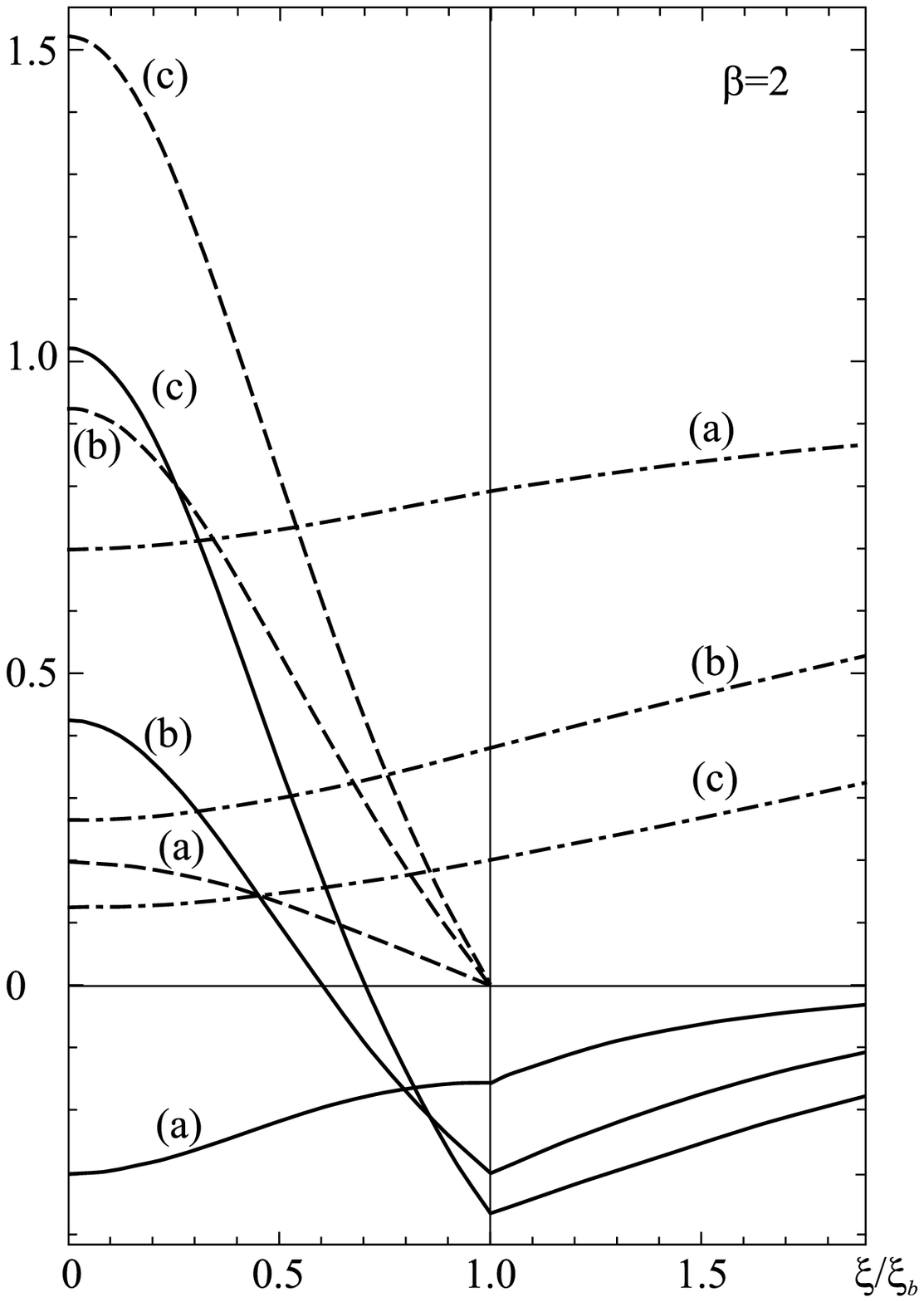}
  \end{center}
\end{minipage}\hfill
\caption{
The metric function $g_{tt}=e^{\nu}$ (dashed-dotted lines),
the total energy density $T_0^0$ (solid lines)
from the right-hand side of Eq.~\eqref{Einstein-00_dmls}
(in units of $\varphi_1^2$), and the fluid energy density $B  (1+\sigma n \theta) \theta^n$
(dashed lines)  are shown
 as functions of the relative radius $\xi/\xi_b$
for the isotropic fluid with $\beta=0$ (left panel) and for the anisotropic fluid  with $\beta=2$
(right panel). Since the solutions are symmetric with respect to $\xi=0$, the graphs are shown only
for $\xi>0$. The curves marked by
(a), (b), and (c) are obtained with the values of the parameter $B$ from Figs.~\ref{fig_sigma}(a), 
\ref{fig_sigma}(b), and \ref{fig_sigma}(c), respectively.  The thin vertical lines correspond to the boundary of the fluid.
For all plots, the characteristic size $L$ is taken as 10 km.
Asymptotically, as $\xi\to\pm\infty$, the spacetime is flat
with $\Sigma\to |\xi|$ and $e^{\nu}\to 1$
from  below.
}
\label{fig_energ_metr}
\end{figure}

We exhibit several examples for neutron-star-plus-wormhole solutions
in Fig.~\ref{fig_energ_metr}.
In particular, we show the metric function $g_{tt}=e^{\nu}$,
the total energy density $T_0^0$,
and the fluid energy density $B  (1+\sigma n \theta) \theta^n$
versus the relative radius $\xi/\xi_b$.
The values of the parameter $B$ are taken from Figs.~\ref{fig_sigma}(a)--\ref{fig_sigma}(c), respectively,
both for an isotropic fluid ($\beta=0$) and an anisotropic fluid ($\beta=2$).

As seen in the figure, the graphs of the total energy density
exhibit a characteristic kink at the boundary of the fluid $\xi=\xi_b$.
This is because the energy density of the fluid is equal to zero at that point,
whereas its derivative differs from zero.
This feature is typical for polytropic fluids.
Beyond the fluid, there exists the scalar field ``tail''
whose energy density goes to zero  as $\xi \to \infty$.
Correspondingly, the spacetime becomes asymptotically flat with $g_{tt}\to 1$.

\begin{figure}[t]
\centering
  \includegraphics[height=9cm]{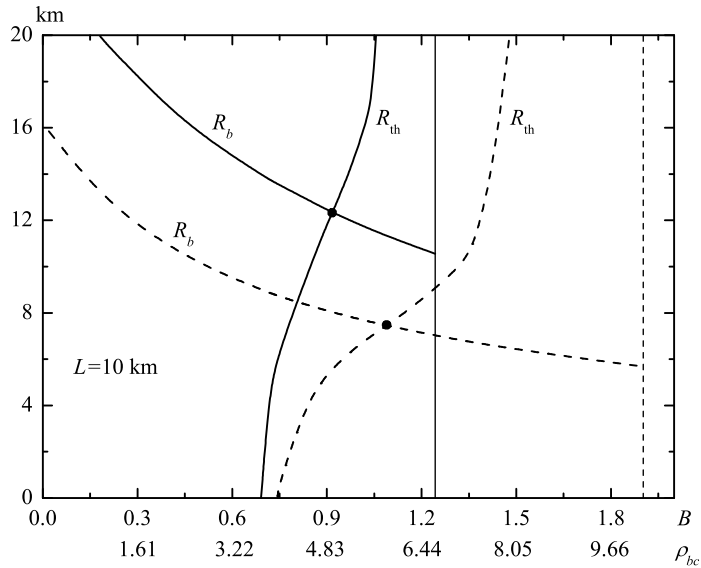}
\vspace{-1.cm}
\caption{The positions of the throat $R_{\text{th}}=L \xi_{\text{th}}$
and of the boundary of the fluid $R_b=L \xi_b$ (both in kilometers)
for the anisotropy parameter $\beta=0$ (solid lines) and $\beta=2$ (dashed lines).
The thin vertical lines correspond to
$B= B_{\text{crit}}$ at which $\Sigma_c\to \infty$.
Above the points of intersection of the curves $R_{\text{th}}$ and $R_b$
(shown by the bold dots), the throats are located beyond the fluid.
}
\label{fig_sizes}
\end{figure}

The positions of the throat $\xi_{\text{th}}$ and of the boundary of the fluid $\xi_b$
shown in Fig.~\ref{fig_sigma}
correspond to the three selected values of $B$ (or equivalently $\rho_{bc}$).
In Fig.~\ref{fig_sizes}
the behavior of these characteristic physical quantities is shown
as a function of $B$.
The numerical calculations indicate that for the values of the fluid parameters
and of the anisotropy parameter $\beta$ used here,
the central value $\Sigma_c\to \infty$ as $B\to B_{\text{crit}}$,
but the size of the throat $\Sigma_{\text{th}}$,
and correspondingly the mass of the throat, remain finite.
In order to understand why this happens,
let us consider the masses associated with the various components
appearing in the system.

\subsection{Masses}

Following Visser \cite{Visser}, we define the total mass
$M$ of the configuration
in terms of an integral at spatial infinity,
representing the Arnowitt-Deser-Misner mass.
Thus $M$ corresponds to the asymptotic mass as measured
by an observer in one of the asymptotically flat regions.
Since we here consider symmetric wormholes,
in both asymptotically flat regions the same value
for the mass is found.
In the following, we therefore consider only the region $r \ge 0$
in detail, keeping in mind that the region $r\le 0$
has identical properties.

For the spherically symmetric metric \eqref{metric_wh_poten},
we consider a volume enclosed by a sphere with
circumferential radius $R_c$, corresponding to the center
of the configuration, and another sphere with
circumferential radius $R>R_c$.
The mass $m(r)$ associated with this volume
can then be defined as follows:
\begin{equation}
\label{mass_dm}
m(r)=\frac{c^2}{2 G}R_c+\frac{4\pi}{c^2}\int_{R_c}^{r} T_0^0 R^2   dR.
\end{equation}
When $B<B_1$, the circumferential radius $R_c$ corresponds to the radius
of the wormhole throat defined by $R_c=\text{min}\{R(r)\}$.
However, when $B_1<B<B_{\text{crit}}$, as mentioned above,
$R_c$ correspond to an equator,
while the two throats are located symmetrically away from the center.

In the dimensionless variables of Eqs.~\eqref{dimless_xi_v} and \eqref{theta_def}
the expression \eqref{mass_dm} takes the form
\begin{equation}
\label{mass_dmls}
m(\xi)=M^*
\left\{\Sigma_c+
\int_{0}^{\xi} \Big[
B  (1+\sigma n \theta)\theta^n-\frac{1}{2}\phi^{\prime 2}
\Big]\Sigma^2 \frac{d\Sigma}{d\xi'}d\xi'
\right\},
\end{equation}
where the coefficient $M^*$ in front of the curly brackets has the dimension of mass
$$
M^*=\frac{c^3}{2}\sqrt{\frac{B}{8\pi G^3 \rho_{b c}}}.
$$

Note that the total mass $M$ is then obtained by taking the upper limit of the integral to infinity,
since the energy density of the scalar field becomes equal to zero only asymptotically,
as $\Sigma\to \infty$.
Note also that in evaluating the above integral it is necessary
to perform the calculations in the internal and external regions separately.

While the asymptotic value $M=\lim\limits_{\xi \to \infty}m(\xi)$
corresponds to the total mass of the configuration,
we would now like to subdivide this expression
into the following four dimensionless components:
\begin{equation}
\label{dim_mass}
M=M^*\left(
{\cal M}_c+ {\cal M}_{\text{fl}}+
{\cal M}_{\text{sfint}}+{\cal M}_{\text{sfext}}\right),
\end{equation}
where we associate
$${\cal M}_c=\Sigma_c$$
with the boundary at the circumferential radius of the
center of the configuration  $\Sigma_c$. Further, we associate
$${\cal M}_{\text{fl}}=\int_{0}^{\xi_b}
B  (1+\sigma n \theta)\theta^n \Sigma^2 \frac{d\Sigma}{d\xi'}d\xi'$$
with the mass of the fluid,
$${\cal M}_{\text{sfint}}=-\frac{1}{2}\int_{0}^{\xi_b}
\phi^{\prime 2} \Sigma^2 \frac{d\Sigma}{d\xi'}d\xi'$$
with the internal part of the mass of the scalar field, and
$${\cal M}_{\text{sfext}}=-\frac{1}{2}\int_{\xi_b}^{\infty}
\phi^{\prime 2} \Sigma^2 \frac{d\Sigma}{d\xi'}d\xi'$$
with the external part of the mass of the scalar field.

For $B<B_1$, the throat is located at the center, and its mass is
${\cal M}_{\text{th}}={\cal M}_c$.
In this case, since $d\Sigma/d\xi>0$,
the expression for ${\cal M}_{\text{fl}}$ is positive,
and it may be interpreted as the total mass of the fluid.
However, for $B>B_1$, we obtain $d\Sigma/d\xi<0$
either in a part of the interval $0<\xi<\xi_b$
where the neutron matter is located, or even in the full interval
$0<\xi<\xi_b$.
Consequently, we here obtain either a negative contribution to the integral for
${\cal M}_{\text{fl}}$, or even a negative value for the full integral.
It is clear that in this case the interpretation of
${\cal M}_{\text{fl}}$ as the total mass of the fluid is problematic.
However, the expression for the total mass of the system
\eqref{dim_mass} does give the correct value.

For this reason, it is interesting to consider the proper mass $M_{\text{prop}}$
of the neutron matter:
\begin{equation}
\label{mass_nm_prop}
M_{\text{prop}}\equiv m_b N=4\pi \int_0^{r_b} \rho_b R^2 dr=
4\pi \rho_{bc}L^3\int_0^{\xi_b} \theta^n \Sigma^2 d\xi .
\end{equation}
$M_{\text{prop}}$ is equal to the mass
which the baryons of the star would possess altogether,
if they were dispersed throughout a volume so large
that all types of interactions between them could be neglected.
Evaluating the expression (\ref{mass_nm_prop}) for the proper mass,
we find the number of neutrons $N$ in the system.

\begin{figure}[t]
\centering
  \includegraphics[height=9.cm]{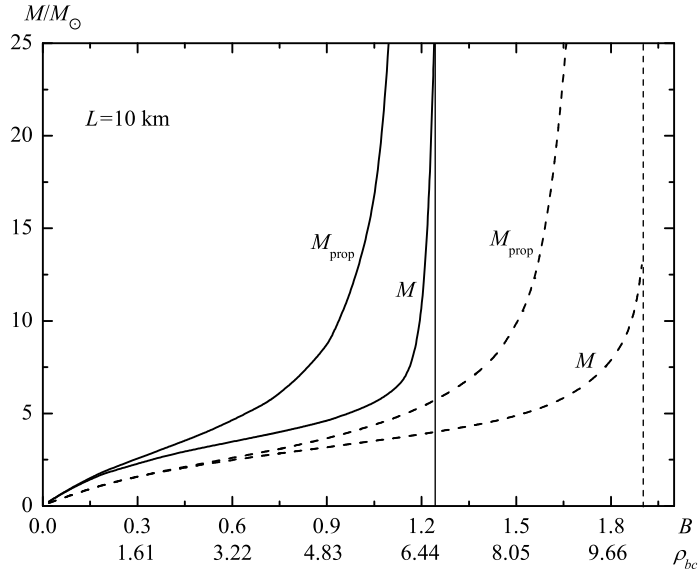}
\vspace{-1.cm}
\caption{The total mass of the configuration
$M$ and the proper mass of the neutron matter $M_{\text{prop}}$ (both in solar mass units)
for the anisotropy parameter $\beta=0$ (solid lines) and $\beta=2$ (dashed lines).
The thin vertical lines correspond to $B= B_{\text{crit}}$.
}
\label{fig_masses_B}
\end{figure}

We exhibit the total mass $M$ \eqref{dim_mass}
and the proper mass of the neutron matter $M_{\text{prop}}$ \eqref{mass_nm_prop}
versus the parameter $B$ in Fig.~\ref{fig_masses_B}.
It is seen from this figure that the total mass $M$ remains finite,
when $B\to B_{\text{crit}}$ .
On the other hand,
the values of $\Sigma_c$ and thus ${\cal M}_c$ diverge,
when $B\to B_{\text{crit}}$.
Clearly, this divergence must be canceled by
another diverging term to yield the observed finite total mass.

Let us introduce the mass of the throat for these configurations:
$${\cal M}_{\text{th}}={\cal M}_c+{\cal M}_{\text{fl}}^{\text{neg}}+{\cal M}_{\text{sfint}}^{\text{neg}}+{\cal M}_{\text{sfext}}^{\text{neg}}.$$
Here the expressions for the masses with the index ``neg''
refer to the values of the corresponding mass integrals
with negative derivative $d\Sigma/d\xi$.
Thus these integrals are evaluated inside the belly region up to
the throat.

When $B\to B_{\text{crit}}$ the integrals
${\cal M}_{\text{sfint}}^{\text{neg}}$ and
${\cal M}_{\text{sfext}}^{\text{neg}}$ also diverge to plus infinity,
and even much stronger than ${\cal M}_c$.
However, at the same time
the integral ${\cal M}_{\text{fl}}^{\text{neg}}$ diverges very strongly
to minus infinity.
Indeed, ${\cal M}_{\text{fl}}^{\text{neg}}$ cancels precisely the divergence
of ${\cal M}_c$, ${\cal M}_{\text{sfint}}^{\text{neg}}$,
and ${\cal M}_{\text{sfext}}^{\text{neg}}$,
to yield a finite value of the mass of the throat ${\cal M}_{\text{th}}$.

The divergence of ${\cal M}_{\text{fl}}^{\text{neg}}$ in turn
is accompanied by a divergence of
the proper mass of the neutron matter, $M_{\text{prop}}\to~\infty$,
and correspondingly by a growth of the number of
neutrons $N\to \infty$.
Thus configurations with $B$ close to $B_{\text{crit}}$
must contain a huge number of particles. 
Since for $B_2<B<B_{\text{crit}}$ the neutron matter is located
completely in the belly region inside the throats,
any light radiated from the star should pass through the throats.
The lensing effects arising in this case are considered in the next subsection.

\subsection{Light passing through the throat}

We now consider the case $B>B_2$, where the two throats are located outside the fluid
[see Fig.~\ref{fig_sigma}(c)].
Thus the neutron matter (the star) is located in the belly region between the two throats,
and any light radiated by the fluid should pass through the throats
to escape to a distant observer.
We would now like to know the intensity
distribution of such radiation.

Following Refs.~\cite{Shatskiy:2004tq,Shatskiy:2007xr,Novikov:2007zz},
we consider light passing through the throat in the equatorial plane,
i.e., in the plane $\theta=\pi/2$ (not to be confused with the fluid density).
In our case the source of the radiation
is the surface of the neutron star located at $\pm \xi_b$
inside the throats, which themselves are located at $\pm \xi_{\text{th}}$;
see Fig.~\ref{fig_sigma}(c).

The path of a light ray in the spherically symmetric gravitational field
described by the metric \eqref{metric_gen}
is determined by the geodesic equation obtained from the Lagrangian
\begin{equation}
-2 {\cal L} = e^\nu c^2 {\dot t}^2 - {\dot r}^2 -
R^2\left( {\dot{\theta}}^2 + \sin^2 \theta\, {\dot{\varphi}}^2 \right)
=0 ,
\label{geodesic1}
\end{equation}
where for geodesics in the equatorial plane $\theta=\pi/2$
and the dot denotes the derivative with respect to an affine parameter.
The cyclic coordinates $t$ and $\varphi$  (not to be confused with $\varphi$
used earlier for the scalar field)
yield the conserved
energy and the conserved angular momentum, proportional to
$\bar E = e^\nu c^2 {\dot t}$ and $\bar \Phi = R^2 \dot{\varphi}$,
respectively.
Insertion of these constants of motion into Eq.~\eqref{geodesic1}
leads to the radial equation
\begin{equation}
{\dot r}^2 = \frac{{\bar{E}^2}}{c^2 e^\nu} - \frac{ {\bar{\Phi}^2}}{R^2}\,.
\label{geodesic2}
\end{equation}
From $\dot{\varphi}$ and ${\dot r}$ we now obtain the dependence of
the angle $\varphi$ on the radial coordinate $r$:
\begin{equation}
\frac{d\varphi}{d r} = \frac{ {\bar{\Phi} }}
{ R^2 \sqrt{ \frac{{\bar{E}^2}}{c^2 e^\nu} - \frac{ {\bar{\Phi}^2}}{R^2} } } .
\end{equation}
Introducing the impact parameter $b= c \Phi/E = c {\bar\Phi}/{\bar E}$
and changing to dimensionless variables,
we obtain for the deflection angle of a photon emitted from the surface of the
star
\begin{equation}
\label{angle_impact}
\delta\varphi=\int_{\xi_b}^\infty \frac{h}{\Sigma^2\sqrt{e^{-\nu}-h^2/\Sigma^2}}\,d\xi,
\end{equation}
where $h=b/L$ is the dimensionless impact parameter.

\begin{figure}[t]
\centering
  \includegraphics[height=8.cm]{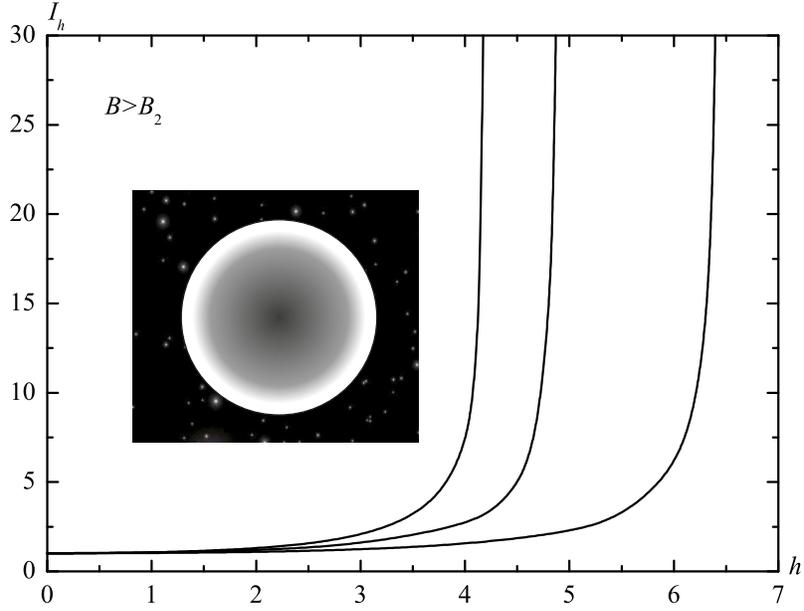}
\caption{The distribution of the intensity $I_h(h)$
(normalized to the corresponding intensity at $h=0$)
of the light passing through the throat
versus the impact parameter $h$.
The curves with
 $B\approx 1.02$
($\rho_{bc}=5.5\times 10^{14} \text{g cm}^{-3}$), $B\approx 1.12$
($\rho_{bc}=6\times 10^{14} \text{g cm}^{-3}$), and $B\approx 1.21$
($\rho_{bc}=6.5\times 10^{14} \text{g cm}^{-3}$) are shown from left to right.
The anisotropy parameter is chosen as $\beta=0$.
The inset shows how a distant observer would see such a configuration.
}
\label{fig_intens}
\end{figure}

Next, following Refs.~\cite{Shatskiy:2007xr,Novikov:2007zz},
we determine the change in the total intensity of the light from different directions
$I_{\text{tot}}$ as a function of the impact parameter:
$$
\frac{d I_{\text{tot}}}{d h}\equiv I_h(h)=\frac{d I_{\text{tot}}}{d\delta\varphi}\frac{d\delta\varphi}{d h}=\text{const} \frac{d\delta\varphi}{d h},
$$
where the intensity density per unit angle $I_{\delta\varphi}\equiv d I_{\text{tot}}/d\delta\varphi$
is taken to be constant, since the intensity
is assumed to be isotropic.
Using Eq.~\eqref{angle_impact}, we find
\begin{equation}
\label{inten_impact}
I_h(h)=\text{const}\int_{\xi_b}^\infty \frac{d\xi}{\Sigma^2 e^{\nu}\left(e^{-\nu}-h^2/\Sigma^2\right)^{3/2}}.
\end{equation}

A distant observer will see light from the neutron star,
whose intensity has a minimum at zero impact parameter
and a maximum for a value of the impact parameter $h_{\text{max}}$,
that is determined by the vanishing of the radicand in Eq.~\eqref{angle_impact}.
Like all other characteristics of the neutron-star-plus-wormhole configurations,
the value of $h_{\text{max}}$ depends ultimately on the value of the parameter $B$.
Figure~\ref{fig_intens} shows the distributions of the intensity
as obtained from Eq.~\eqref{inten_impact} for different values of
$B>B_2$.

The numerical calculations indicate that, as $B$ (and hence the central density of the fluid)
increases, the size of the throat increases,
which implies a growing impact parameter $h_{\text{max}}$.
The presence of such an effect in gravitational lensing
is visible to a distant observer,
who will see a radiating object
in the form of a ring of light with sharp external boundaries and diffuse internal boundaries
(see Fig.~\ref{fig_intens}, and also Fig.~3 in Ref.~\cite{Novikov:2007zz}).

\subsection{Tidal accelerations}

Let us now address the tidal accelerations in the gravitational field
of the neutron-star-plus-wormhole configurations.
Any two separate points of a body embedded in an inhomogeneous gravitational field
are subject to slightly different accelerations. This results in the appearance of a tidal force.
We here estimate the tidal accelerations for the neutron-star-plus-wormhole configurations.

Following Ref.~\cite{Visser}, we consider the radial and transverse components
of the tidal acceleration,
which in the metric~\eqref{metric_wh_poten} are given by
\begin{eqnarray}
\label{tid_accel_rad}
&&\frac{1}{c^2}(\Delta a)_{||}=R_{\hat{0}\hat{1}\hat{0}\hat{1}}(\Delta x)_{||}=
-\frac{1}{2}\left(\nu^{\prime\prime}+\frac{1}{2}\nu^{\prime 2}\right)(\Delta x)_{||}~,\\
\label{tid_accel_tan}
&&\frac{1}{c^2}(\Delta a)_{\bot}=\frac{1}{1-(v/c)^2}
\left[R_{\hat{2}\hat{0}\hat{2}\hat{0}}+(v/c)^2 R_{\hat{2}\hat{1}\hat{2}\hat{1}}\right](\Delta x)_{\bot}
=
\frac{1}{1-(v/c)^2}\left[-\frac{1}{2}\nu^\prime\frac{R^\prime}{R}+(v/c)^2\frac{R^{\prime\prime}}{R}\right](\Delta x)_{\bot}
\end{eqnarray}
[see Eqs.~(13.4) and (13.6) of Ref.~\cite{Visser},
here rewritten with our signature of the metric].
The hats on the indices of the Riemann tensor indicate the use of an orthonormal frame;
$(\Delta x)_{||}$ and $(\Delta x)_{\bot}$
correspond to distances between two points of the body
in the radial and transverse directions, respectively;
$v$ is the three-dimensional radial velocity of the body.

Since we here consider only static configurations,
for which the neutron matter is at rest,  we set
$v=0$.
Then Eqs.~\eqref{tid_accel_rad} and \eqref{tid_accel_tan}
can be rewritten in the dimensionless variables of Eq.~\eqref{dimless_xi_v}
as follows:
\begin{eqnarray}
\label{tid_accel_rad_dmls}
&&(\Delta a)_{||}=
-\frac{c^2}{2 L}\left(\nu^{\prime\prime}+\frac{1}{2}\nu^{\prime 2}\right)(\overline{\Delta x})_{||}~,\\
\label{tid_accel_tan_dmls}
&&(\Delta a)_{\bot}=
-\frac{c^2}{2 L}\nu^\prime\frac{\Sigma^\prime}{\Sigma}(\overline{\Delta x})_{\bot},
\end{eqnarray}
where the bar on $\Delta x$ denotes the dimensionless quantity.

\begin{figure}[t]
\begin{minipage}[t]{.49\linewidth}
  \begin{center}
  \includegraphics[width=8cm]{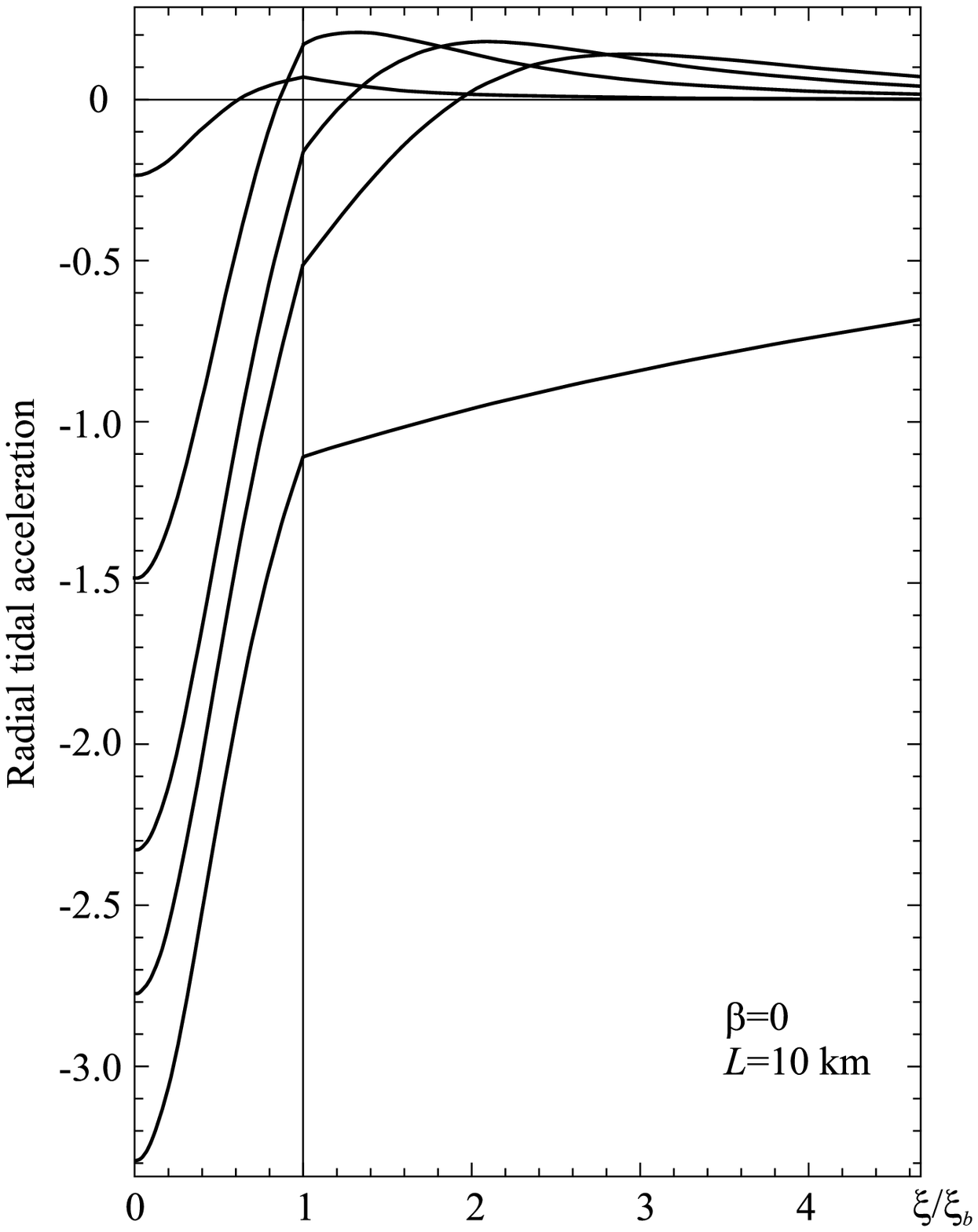}
  \end{center}
\end{minipage}\hfill
\begin{minipage}[t]{.49\linewidth}
  \begin{center}
  \includegraphics[width=8cm]{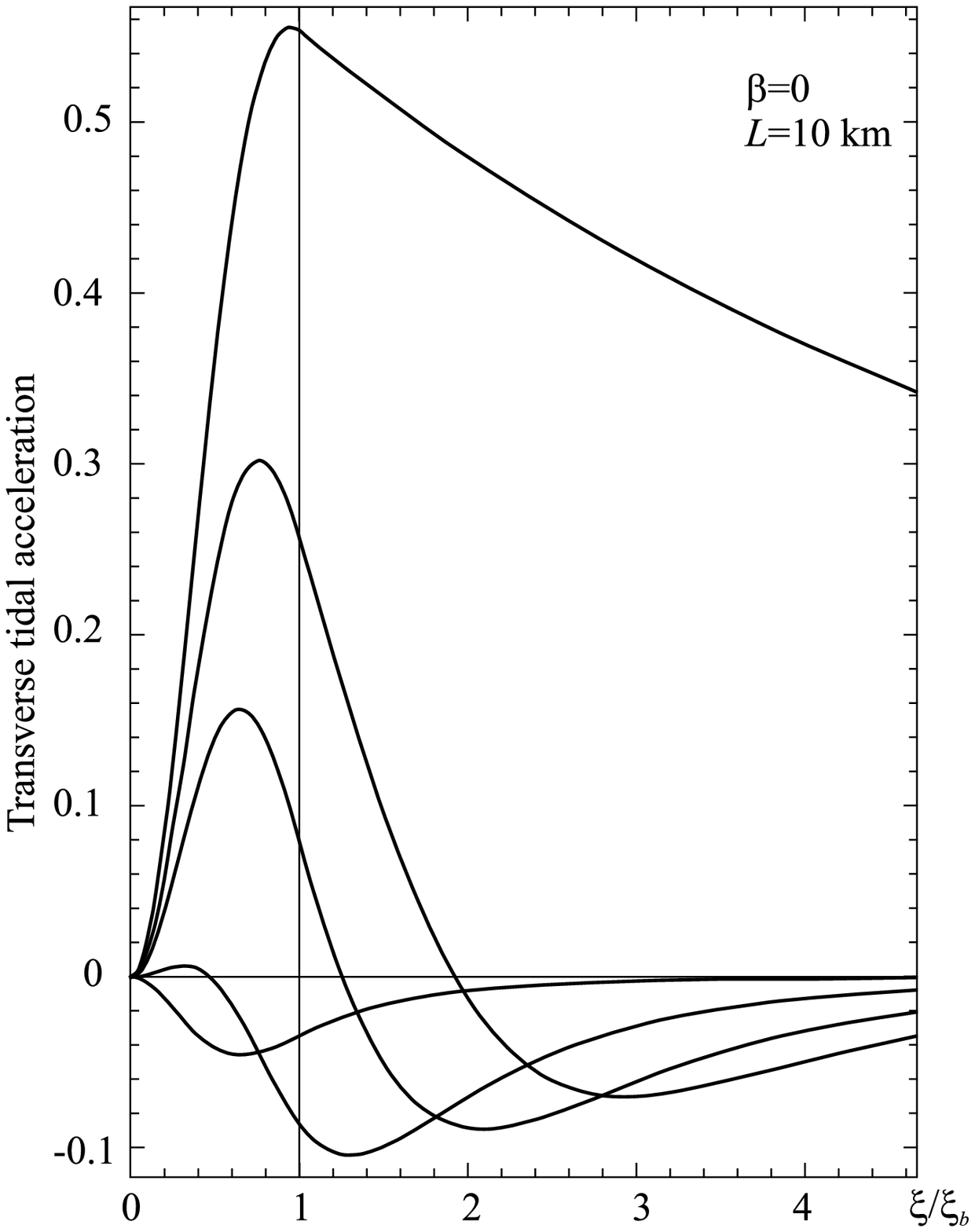}
  \end{center}
\end{minipage}\hfill
\caption{
The relative radial tidal acceleration from Eq.~\eqref{rel_tid_accel_rad} (left panel)
and the relative transverse tidal acceleration from Eq.~\eqref{rel_tid_accel_tan} (right panel)
are shown versus the relative radius $\xi/\xi_b$.
In the left panel the curves correspond to
$\rho_{bc}=6.6754$ ($B\approx B_{\text{crit}}\approx 1.243$),
$\rho_{bc}=6$ ($B\approx 1.12$), $\rho_{bc}\approx 5.37$ ($B= 1$),
$\rho_{bc}= 4$ ($B\approx 0.75$), and
$\rho_{bc}= 1$ ($B\approx 0.19$)
(all values of $\rho_{bc}$ are in units of $10^{14} \text{g cm}^{-3}$), from bottom to top.
(For the corresponding distributions of the neutron matter, see Fig.~\ref{fig_sigma}.)
In the right panel the same values of $\rho_{bc}$ are used,
but the order of the curves is reversed
(the top and the bottom curves correspond,
respectively, to the largest and to the smallest $\rho_{bc}$).
The negative and positive values of the accelerations correspond to
compressive and stretching tidal forces, respectively.
The thin vertical lines indicate the boundary of the fluid.
Asymptotically the accelerations go to zero like $\pm \xi^{-3}$,
where the plus (minus) sign corresponds to the radial (transverse) tidal accelerations.
}
\label{fig_tid_accel}
\end{figure}

Let us now compare the tidal accelerations of the following two systems:
(i) the neutron-star-plus-wormhole configuration considered here
and (ii) an ordinary neutron star
modeled by the same EOS \eqref{eqs_NS_WH}.
For simplicity, let us consider only
the case where the neutron fluid is isotropic,
i.e., $\beta=\alpha=0$.
The central values of the tidal accelerations will then be

\noindent
(i) for the mixed star-plus-wormhole system
\begin{eqnarray}
\label{tid_accel_rad_centr_mixed}
&&(\Delta a)_{||}|_c=-\frac{c^2}{2 L}B\left[1+\sigma(n+3)\right](\overline{\Delta x})_{||},\\
\label{tid_accel_tan_centr_mixed}
&&(\Delta a)_{\bot}|_c=0;
\end{eqnarray}
(ii) for the neutron star
\begin{eqnarray}
\label{tid_accel_rad_centr_NS}
&&(\Delta a)_{||}|_c=-\frac{c^2}{2 L}\left[1/2+\sigma(1+ n/2)\right](\overline{\Delta x})_{||},\\
\label{tid_accel_tan_centr_NS}
&&(\Delta a)_{\bot}|_c=-\frac{c^2}{2 L}\left[1/2+\sigma(1+ n/2)\right](\overline{\Delta x})_{\bot}.
\end{eqnarray}
Thus the neutron star has the same values for both components of the tidal acceleration.

To compare the two systems, it is convenient to use the same scale for the characteristic length $L$.
Since in considering
the neutron-star-plus-wormhole systems we used $L=10\, \text{km}$,
we take the same~$L$ for the neutron star.
This choice corresponds to a neutron star with a
central density $\rho_{b c}\approx 5.37 \times 10^{14} \text{g cm}^{-3}$
and a total mass $M \approx 3.16 M_\odot$,
close to the maximum mass of a neutron star for such an EOS
(cf.~Fig. 3 in Ref.~\cite{Dzhunushaliev:2012ke}).

Since for such neutron stars the maximum (modulus) of the tidal accelerations will
occur at the center, it is convenient to normalize
the tidal accelerations of our mixed configurations with respect to these central values.
We then obtain
\begin{eqnarray}
\label{rel_tid_accel_rad}
\text{(relative radial tidal acceleration)} &=&
\frac{[\text{Eq.}~\eqref{tid_accel_rad_dmls}\, \text{for the mixed system}]}{[\text{the modulus of Eq.}~\eqref{tid_accel_rad_centr_NS}]}=
-\frac{\nu^{\prime\prime}+\nu^{\prime 2}/2}{1/2+\sigma(1+ n/2)},\\
\label{rel_tid_accel_tan}
\text{(relative transverse tidal acceleration)} &=&
\frac{[\text{Eq.}~\eqref{tid_accel_tan_dmls}\, \text{for the mixed system}]}{[\text{the modulus of Eq.}~\eqref{tid_accel_tan_centr_NS}]}=
-\frac{\nu^\prime \Sigma^\prime/\Sigma}{1/2+\sigma(1+ n/2)}.
\end{eqnarray}

The results of the numerical calculations are shown in Fig.~\ref{fig_tid_accel}.
Indeed, the maximum (modulus) of the tidal accelerations occurs in the central regions of our mixed configurations.
They are comparable in size
to those of the neutron stars used here for comparison.
Even in the case when $B\to  B_{\text{crit}}$,
the radial tidal acceleration is by only a factor of approximately
$3.3$ larger than that of an ordinary neutron star.

\section{Linear stability analysis}
\label{linear_stab}

We now consider spherically symmetric perturbations of
the above equilibrium configurations. In our previous work
 \cite{Dzhunushaliev:2013lna} we performed the
linear stability analysis for the case of an isotropic neutron fluid and a scalar field with
a  quartic potential.
One can find there a detailed derivation of the corresponding equations for the perturbations.
Therefore we here simply employ the equations from Ref.~\cite{Dzhunushaliev:2013lna},
except for a small change necessary to incorporate the anisotropy of the fluid.

It is convenient to use the general form of metric
\eqref{metric_gen}, in which
the components of the four-velocity of the fluid
can be written as follows \cite{Chandrasekhar:1964zz}:
$$
u^0=e^{-\nu_0/2}, \quad u_0=e^{\nu_0/2}, \quad u^1=e^{-\nu_0/2} v, \quad u_1=-e^{\lambda_0-\nu_0/2} v,
$$
with the three-velocity
$$
v=\frac{d r}{d x^0} \ll 1 ~.
$$
The index 0 on the metric functions indicates the static, zeroth-order solutions of the Einstein equations.
Then the functions $\nu$, $\lambda$, $\mu$, $\varepsilon$, $p$, and $\varphi$
appearing in the system can be presented in the harmonic form
\begin{equation}
\label{perturbations}
y=y_0+y_p(\xi) e^{i\omega x^0}~,
\end{equation}
where $y$ denotes any one of the above functions,
$y_p(\xi)$ depends only on the spatial coordinate $\xi$,
the index $p$ indicates the perturbation, and $\omega$ is the frequency of the radial oscillations.

Next, using  the gauge choice $\lambda_0=0$ and $\nu_p=\lambda_p-2\mu_p$, one can derive the following set of perturbed equations
(for details, see Ref.~\cite{Dzhunushaliev:2013lna}): the scalar field equation
 \begin{equation}
\label{phi_pert_gen_lam0}
\phi_p^{\prime\prime}+
\frac{1}{2}\left(\nu_0^\prime+2\mu_0^\prime\right)\phi_p^\prime
+\omega^2 e^{-\nu_0}\phi_p=0,
\end{equation}
the perturbed (0-0) and (2-2) components of the Einstein equations,
\begin{eqnarray}
\label{Einstein-00pert_lam0}
&&\mu_p^{\prime\prime}+\frac{1}{2}\mu_0^\prime\left(3\mu_p^\prime-\lambda_p^\prime\right)
-\left(\mu_0^{\prime\prime}+\frac{3}{4}\mu_0^{\prime 2}\right)\lambda_p
+e^{-\mu_0}\mu_p
\nonumber \\
&&=
-n B \left[\frac{1}{\theta_0}+\sigma(n+1)\right]\theta_0^n\theta_p+\phi_0^\prime\left(\phi_p^\prime-\frac{1}{2}\phi_0^\prime \lambda_p\right),
\\
\label{Einstein-22pert_lam0}
&&\lambda_p^{\prime\prime}-\mu_p^{\prime\prime}
+\frac{1}{2}\nu_0^\prime\left(\lambda_p^\prime-3\mu_p^\prime\right)
-\lambda_p\left[
\mu_0^{\prime\prime}+\nu_0^{\prime\prime}+\frac{1}{2}\left(\mu_0^{\prime 2}+\nu_0^{\prime 2}+\mu_0^\prime \nu_0^\prime\right)
\right]
+\omega^2e^{-\nu_0}(\mu_p+\lambda_p) \nonumber
\\
&&=
2\left[B\sigma(n+1)\left(1+\frac{\alpha}{2}\right)
\theta_0^n \theta_p+\phi_0^\prime\left(\phi_p^\prime-\frac{1}{2}\phi_0^\prime \lambda_p\right)
\right],
\end{eqnarray}
and, finally, the equation which follows from the $i=1$ component of
the law of conservation of energy and momentum, $T^k_{i; k}=0$,
\begin{eqnarray}
\label{conserv_osc_pert_mex}
&&\frac{1}{2}\omega^2 e^{-\nu_0}\left[
2\mu_p^\prime-\mu_0^\prime\lambda_p+\left(\mu_0^\prime-\nu_0^\prime\right)\mu_p
\right]-
B\sigma(n+1)(1-\alpha)\frac{d}{d\xi}\Big(\theta_0^n \theta_p\Big)
\nonumber
 \\
&&+\phi_0^{\prime\prime}\left(\phi_p^\prime-\frac{1}{2}\phi_0^\prime \lambda_p\right)
+\phi_0^\prime\left[\phi_p^{\prime\prime}-\frac{1}{2}\left(\phi_0^{\prime\prime}\lambda_p+
\phi_0^\prime \lambda_p^\prime\right)\right]
-\frac{1}{2}B \theta_0^n\left[\frac{n}{\theta_0}+\sigma(n+1)(n+1-\alpha)\right]\theta_p\nu_0^\prime \nonumber
\\
&&-\frac{1}{2}B \theta_0^n\left[1+\sigma(n+1-\alpha)\theta_0\right](\lambda_p^\prime-2\mu_p^\prime)
+\frac{1}{2}\phi_0^{\prime 2}(\lambda_p^\prime-2\mu_p^\prime)
+\nu_0^\prime\phi_0^\prime\left(\phi_p^\prime-\frac{1}{2}\phi_0^\prime \lambda_p\right)\nonumber
\\
&&+
\mu_0^\prime\left[\frac{3}{2}\alpha B \sigma(n+1)\theta_0^n\theta_p
+2\phi_0^\prime\left(\phi_p^\prime-\frac{1}{2}\phi_0^\prime \lambda_p\right)\right]
+\mu_p^\prime\left(\frac{3}{2}\alpha B \sigma\theta_0^{n+1}+\phi_0^{\prime 2}\right)=0.
\end{eqnarray}

Thus, for the four functions $\phi_p$, $\lambda_p$, $\mu_p$, $\theta_p$,
we have the set of four equations
\eqref{phi_pert_gen_lam0}-\eqref{conserv_osc_pert_mex}
to investigate the stability of the configurations.
For this set of equations, we choose the following boundary conditions
at $\xi=0$:
\begin{equation}
\label{bound_cond_pert_mex}
\lambda_p(0)=\lambda_{p 0}, \quad
\mu_p(0)=\mu_{p 0},\quad
\theta_p(0)=\theta_{p 0}, \quad
\phi_p(0)=0,\quad \phi_p^\prime(0)=\phi_{p 1},
\end{equation}
where $\lambda_p$, $\mu_p$, and $\theta_p$ are even functions,
while $\phi_p$ is an odd function. The value of $\phi_{p 1}$
can be found from the perturbed (1-1) component of the Einstein equations in the following form:
$$
\phi_{p 1}=B\sigma (n+1)(1-\alpha)\theta_{p 0}+\frac{1}{2}\lambda_{p 0}-\left(\omega^2 e^{-\nu_c}+\frac{1}{\Sigma_c^2}\right)\mu_{p 0}.
$$

\begin{figure}[t]
\centering
  \includegraphics[height=9.cm]{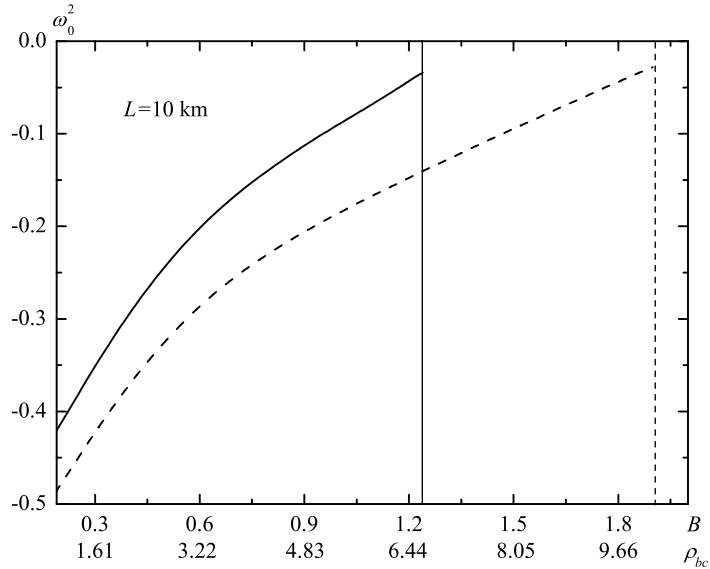}
\vspace{-1.cm}
\caption{The  lowest eigenvalue $\omega_0^2$ is shown as a function of $B$
for the anisotropy parameter $\beta=0$ (solid line) and $\beta=2$ (dashed line).
The thin vertical lines correspond to $B=B_{\text{crit}}$.
}
\label{fig_omega_B}
\end{figure}

Thus the system contains three free parameters:
$\lambda_{p 0}$, $\mu_{p 0}$, and $\theta_{p 0}$.
Their values are chosen such that the following conditions are satisfied.
(i) At the boundary of the fluid, $\xi=\xi_b$,
the value of $\theta_p$ should remain finite to ensure
that $p_p=K (n+1) \rho_{b c}^{1+1/n}\theta_0^n \theta_p$  meets the condition $p_p=0$
at the boundary where $\theta_0=0$
 [see, e.g., Eq.~(60) in Ref.~\cite{Chandrasekhar:1964zz}].
(ii) Asymptotically,
as $\xi\to  \infty$, the perturbations $\lambda_p$, $\mu_p$, and $\phi_p$ should tend to zero.
In this connection
it is useful to determine the asymptotic behavior of the solutions.
This can be given in analytic form.

\medskip
\noindent (A) {\it Static solutions}.--
\medskip

 $\left\{  \begin{tabular}{l}
$\phi_0\to C_1-C_2 /\xi$;\\[\medskipamount]
$\Sigma_0 \to \xi$, \quad $\Sigma_0^\prime \to 1-C_3/\xi$;\\[\medskipamount]
$e^{\nu_0}\to 1-2\, C_3/\xi$.\\[\medskipamount]
\end{tabular}  \right.  $

\medskip
\noindent (B) {\it Perturbations}.--
\medskip

$\left\{ \begin{tabular}{l}
$\phi_p\to C_4 \exp{\left(-\sqrt{-\omega^2}\xi\right)}\Big/\xi$;\\[\medskipamount]
$\mu_p \to C_5  \exp{\left(-\sqrt{-\omega^2}\xi\right)}$;\\[\medskipamount]
$\lambda_p \to -C_5  \sqrt{-\omega^2}\xi \exp{\left(-\sqrt{-\omega^2}\xi\right)}$.\\[\medskipamount]
\end{tabular}\right.  $

\noindent Here  the $C_i$ are integration constants.
Note that in the above expressions the frequency $\omega^2$ carries a minus sign
under the square root.
Therefore,
to obtain decaying solutions for the perturbations,  $\omega^2$ should be negative.
If this were not the case, the perturbations would be oscillating along the radius.
In such a case
the derivative of the scalar field perturbation $\phi_p^\prime$
could become asymptotically larger than the static solution
$\phi_0^\prime$,
which would be in contradiction to the essence of the perturbation method.
Thus the perturbation method employed
here works only for negative  $\omega^2$.

Let us now use Eqs.~\eqref{phi_pert_gen_lam0}-\eqref{conserv_osc_pert_mex}
together with the boundary conditions
\eqref{bound_cond_pert_mex} to find the eigenvalue $\omega^2$.
The question of stability is thus reduced to a study of the possible
values of $\omega^2$.
If any of the values of $\omega^2$ are found to be negative,
then the perturbations will grow and the
configurations in question will be unstable against radial oscillations.

The results of the calculation of the lowest eigenvalue
$\omega_0^2$ are shown in Fig.~\ref{fig_omega_B},
where $\omega_0^2$ is presented as a function of the parameter $B$.
As background solutions we employ the static solutions obtained in Sec.~\ref{num_calc}.
The initial value  $\mu_{p}(0)$ in Eq.~\eqref{bound_cond_pert_mex}
is chosen to be $\mu_{p 0}=1$,
and the values $\lambda_{p 0}$ and $\theta_{p 0}$ are chosen in such a way
that the solutions exhibit the asymptotic behavior shown in (B).

It is seen from Fig.~\ref{fig_omega_B} that
the square of the eigenfrequency remains always negative, independent of $B$.
One might naively expect that the inclusion of an anisotropy of the fluid,
which allows one to increase the central fraction of the fluid in the system
(that provides the possibility of obtaining solutions with larger $B$),
would favor the stabilization of the solutions.
This does not happen, however,
and $\omega_0^2$ remains always negative up to the critical values $B_{\text{crit}}$.
Thus, the configurations under consideration are always unstable against linear perturbations.

\section{Conclusion}
\label{conclusion}

In the present paper we have considered neutron-star-plus-wormhole systems
in which a wormhole,
supported by a massless ghost scalar field,
is threaded by ordinary (neutron) matter.
In contrast to the configurations considered in Ref.~\cite{Dzhunushaliev:2013lna},
we have here extended those studies to the case
where the central densities of the scalar field and the neutron fluid are comparable,
i.e., where the parameter $B$ is large, $B\sim 1$.
This has allowed us to obtain systems with double-throat wormholes.
Also, for a more realistic modeling of the neutron matter at high densities,
we have employed an anisotropic equation of state for the neutron matter,
where the radial and tangential pressures of the fluid are not equal.

Our main results are the following:
\begin{enumerate}
\itemsep=-0.2pt
\item[(1)] 
There exist static regular asymptotically flat solutions
describing neutron-star-plus-wormhole systems
in which the neutron matter is concentrated in a finite-size region.
In the simplest case such configurations may be regarded as
consisting of a neutron star with a wormhole at its core,
with the neutron matter filling the wormhole throat.
For these systems, the parameter $B$ is small,
and the throat is located at the center of the system
($\xi_{\text{th}}=0$); see Fig.~\ref{fig_sigma}(a).
For larger  values of the parameter $B$
double-throat wormholes arise,
where the throats are either still lying within the fluid
[see Fig.~\ref{fig_sigma}(b)]
or where they are located outside the fluid [see Fig.~\ref{fig_sigma}(c)].
\item[(2)]
In the latter case,  presented in Fig.~\ref{fig_sigma}(c), the neutron matter
is completely hidden inside the belly region between the throats.
When the neutron matter radiates light passing through the throats,
this is subjected to gravitational lensing.
This leads to a characteristic intensity distribution (see Fig.~\ref{fig_intens}),
where the apparent brightness increases from the center to the limb of the star.
Note that the distribution of the intensity of the light passing through
the throat differs from the one obtained when considering the case where
radiation does not pass through a throat (see, e.g., Fig.~2 from Ref.~\cite{Bambi:2013nla}).
In principle, such an effect could be observed
by instruments with sufficiently high resolution.
\item[(3)]
The tidal accelerations present in the neutron-star-plus-wormhole systems
are comparable to those of neutron stars
modeled by the same EOS \eqref{eqs_NS_WH}.
From this point of view the neutron-star-plus-wormhole configurations appear to be viable.
\item[(4)]
According to the linear stability analysis of Sec.~\ref{linear_stab},
the square of the lowest eigenfrequency of the perturbations is negative.
This indicates that the neutron-star-plus-wormhole configurations are unstable.
This holds independent of whether the fluid is isotropic or anisotropic.
\end{enumerate}

One might expect that in order to obtain stable neutron-star-plus-wormhole systems
one should start from stable wormholes (see, e.g., Refs.~\cite{Kanti:2011jz,Kanti:2011yv,Bronnikov:2013coa}).
Static wormhole configurations obtained from massless ghost scalar fields
are known to be unstable with respect to linear \cite{Gonzalez:2008wd,Bronnikov:2011if}
and nonlinear perturbations \cite{Gonzalez:2008xk}.
Nevertheless, in this case stabilization
of the wormhole solutions might possibly be achieved by including rotation into the system,
as we recently showed for
rapid rotation of five-dimensional wormholes \cite{Dzhunushaliev:2013jja}.
Thus one might expect that a rapid rotation might also stabilize
the neutron-star-plus-wormhole systems in four dimensions.
This question should be considered in our future studies.

\section*{Acknowledgements}

We gratefully acknowledge support provided by the Volkswagen Foundation.
This work was partially supported by  Grant No.~378 in fundamental research in natural sciences
by the Ministry of Education and Science of Kazakhstan
and by the DFG Research Training Group 1620 ``Models of Gravity.''

\end{document}